\documentclass[amsmath,amssymb,,showpacs,twocolumn,10pt]{revtex4}

\usepackage{psfig}
\usepackage{graphicx}
\usepackage{dcolumn}
\usepackage{bm}
\newcommand{\s}{\ensuremath{\psi(t,r)}}
\newcommand{\n}{\ensuremath{\nu(t,r)}}
\newcommand{\T}{\ensuremath{\theta}}
\newcommand{\pt}{\ensuremath{p_\theta}}
\newcommand{\pr}{\ensuremath{p_r}}
\newcommand{\e}{equation\;} 

\newcommand{\prz}{\ensuremath{p_{r_0}}}

\newcommand{\ptz}{\ensuremath{p_{\theta_0}}}

\begin{document}

\title{The critical role of shear in gravitational collapse}

\author{Pankaj S. Joshi$^*$, Rituparno Goswami$^*$ and Naresh 
Dadhich$^\dag$}

\address{~}

\address{$^*$Tata Institute for Fundamental Research, Mumbai,
India}
\address{$^\dag$Inter-University Center for Astronomy \&
Astrophysics, Pune, India}

\begin{abstract} 
We investigate here how the shearing effects 
present within a collapsing matter cloud influence the outcome of  
gravitational collapse in terms of formation of either a black hole 
or a naked singularity as the final end state. For 
collapse of practically all physically reasonable matter 
fields, we prove that it would always end up in a black hole if it 
is either shear-free or has homogeneous density. Thus it follows that 
whenever a naked singularity forms as end product, the collapsing cloud 
must necessarily be shearing with inhomogeneous density.
Our consideration brings out the physical forces at work which
could cause a naked singularity to result as collapse end state,
rather than a black hole.

\end{abstract}
\maketitle

One of the most intriguing phenomena in black hole physics 
and gravitation theory that has attracted considerable attention in 
recent years is the occurrence of naked singularities as possible 
end state of a continual gravitational collapse. It turns out 
that depending on the nature of the initial data of the
matter distribution and the metric functions, either a
black hole or a naked singularity results as the final outcome 
of an endless collapse (see e.g. 
\cite{rev} 
and references there in
for further details). In fact, the occurrence of naked singularities 
could be generic because the initial data evolving to a naked 
singularity would form an open set 
\cite{gary}. 
Such a scenario 
will be of physical interest because a visible extreme strong gravity
region
could possibly provide us with an opportunity to observe directly the 
effects of 
quantum gravity, as generated by the strong curvature regions.

An obvious question of considerable interest and significance 
is then, what are the possible physical factors which distinguish one 
outcome from the other. In other words, one would like to inquire
what are the physical agencies operating during the process of a 
continual collapse of a massive matter cloud that lead to the formation
of a naked singularity as the end state rather than a black hole, 
or vice versa. Such an investigation should help us understand 
much better the physics of black holes and naked singularity 
formation in 
gravitational collapse. Towards this end, it was recently shown 
\cite{jdm}
that for the spherical
dust collapse the shearing effects and inhomogeneity present 
within the 
collapsing cloud do play a crucial role in delaying the formation 
of the trapped surfaces and the apparent horizon. These could in fact
cause sufficient distortion in the 
geometry of apparent and event horizons, thus exposing the 
singularity to external observers.

Our purpose here is to show that spacetime
shear plays a crucial role towards determining the black hole or naked
singularity as end states of gravitational collapse
for almost all physically reasonable matter fields. We consider 
here collapse for {\it type I} 
matter fields, which is a rather broad class comprising of practically 
{\it all} physically interesting matter forms such as dust, perfect 
fluids, massless scalar fields, and such other forms of matter 
\cite{haw}. 
We show that the collapse always ends in a black hole 
whenever there is
no shear present in the collapsing cloud. In other words, 
whenever a naked singularity has developed as the end state of  
collapse, it is then always due to the distortion of the trapped 
surface geometry caused by the presence of shear. As the 
matter considered here is generic, with non-vanishing radial 
as well as tangential pressures, we argue that generically it is the 
spacetime shear which is the physical agency 
that causes naked singularity as the collapse outcome, as opposed 
to a black hole.

Consider a spherical collapsing 
cloud which can be described by the general metric in the comoving   
coordinates $(t,r,\theta,\phi)$ as given by,
\begin{equation}
ds^2=-e^{2\n}dt^2+e^{2\s}dr^2+R^2(t,r)d\Omega^2
\label{eq:metric}
\end{equation}
The energy-momentum tensor for any matter field which is type I 
 is then given in a diagonal form,  
\begin{equation}
T^t_t=-\rho;\; T^r_r=p_r;\; T^\T_\T=T^\phi_\phi=p_\T
\label{eq:setensor}
\end{equation}
The quantities $\rho$, $p_r$ and $p_\T$ are the energy density, radial 
and tangential pressures respectively. 
We take the matter field 
to satisfy the {\it weak energy condition}, i.e. the energy density 
measured by any local observer be non-negative, and so for any 
timelike vector $V^i$, we must have,
\begin{equation}
T_{ik}V^iV^k\ge0
\end{equation}
which amounts to,
\begin{equation}
\rho\ge0;\; \rho+p_r\ge0;\; \rho+p_\T\ge0
\end{equation}
It should be noted that we are not confining ourselves here 
to any special
type of matter such as e.g. dust or a perfect fluid form, but {\it all}
forms of matter are included where the stress-energy tensor admits
one timelike and three spacelike eigen vectors. In that sense, our
conclusions are rather generic, applying to a large variety of collapse
models.

Now for the metric (\ref{eq:metric}) the Einstein equations 
take the form, in the units $(8\pi G=c=1)$

\begin{eqnarray}
\rho=\frac{F'}{R^2R'}; && p_r=-\frac{\dot{F}}{R^2\dot{R}}
\label{eq:ein1}
\end{eqnarray}
\begin{equation}
\nu'=\frac{2(\pt-p_r)}{\rho+p_r}\frac{R'}{R}-\frac{p_r'}{\rho+p_r}
\label{eq:ein2}
\end{equation}
\begin{equation}
-2\dot{R}'+R'\frac{\dot{G}}{G}+\dot{R}\frac{H'}{H}=0
\label{eq:ein3}
\end{equation}
\begin{equation}
G-H=1-\frac{F}{R}
\label{eq:ein4}
\end{equation}
where,
\begin{eqnarray}
G(t,r)=e^{-2\psi}(R')^2; && H(t,r)=e^{-2\nu}(\dot{R})^2
\end{eqnarray}

In the above, we have introduced the arbitrary function $F=F(t,r)$,
which has an interpretation of
the mass function for the cloud, giving the total mass in a 
shell of comoving radius 
$r$. We have $F\ge0$ from the energy conditions.

The shear tensor for the collapsing matter is given by
\cite{mac},
\begin{equation}
\sigma_\phi^\phi=\sigma_\T^\T=-\frac{1}{2}\sigma_r^r=\frac{1}{3}
e^{-\nu}\left(\frac{\dot{R}}{R}-\dot{\psi}\right)
\end{equation}

Now let us consider the situation when the matter shear vanishes 
identically. This means,
\begin{equation}
\frac{\dot{R}}{R}=\dot{\psi}
\label{eq:psidot}
\end{equation}
or,
\begin{equation}
R=q(r)e^{\psi}
\end{equation}
We can use the scaling freedom available in rescaling the 
radial coordinate $r$, and with a suitable 
rescaling we can always choose $q(r)=r$. 
We then have,
\begin{equation} 
R=re^{\psi}
\label{eq:R}
\end{equation}
It then follows that the spacetime geometry (1) becomes,
\begin{equation}
ds^2=-e^{2\n}dt^2+e^{2\s}\left[dr^2+r^2d\Omega^2\right]
\end{equation}

Hence we see that there are now five total field equations with 
six unknowns as given by, $\rho$, $p_r$, $\pt$, $\psi$, $\nu$ and $F$, 
thus giving us 
the freedom of choice of one free function if we are to complete 
the solution. Also we require the regularity of the initial data, 
and in particular the 
density distribution, at the initial 
surface $t=t_i$ from which the collapse develops.
The collapse condition requires $\dot{R}<0$, which amounts 
to $\dot{\psi}<0$. The singularity occurs at $R(r,t_s(r))=0$,
which implies that in the limit as $t \to t_s$,
\begin{equation}
\psi(r,t_s(r))\rightarrow -\infty
\label{eq:psi}
\end{equation}

It is now possible to integrate the equation (7), using equation 
(\ref{eq:R}), to obtain
\begin{equation}
e^{\n}=a(t)\dot{\psi}
\label{eq:nu}
\end{equation}
where $a(t)$ is an arbitrary function of integration. 
Since the left hand side of 
the above equation is positive by definition, it follows 
that $\dot{\psi}<0$ implies 
$a(t)<0$.

Now let us consider the expansion $\Theta$, for the 
infalling matter congruence. For the metric (\ref{eq:metric}),
the expansion is given as,
\begin{equation}
\Theta(t,r)=\frac{1}{e^\nu}\left[\dot{\psi}+2\frac{\dot{R}}{R}\right]
\label{eq:expansion}
\end{equation}
From \e(\ref{eq:psidot}) and (\ref{eq:nu}) then it can be
easily seen that, in case of a shear free collapse,
\begin{equation}
\Theta(t,r)=\Theta(t)=\frac{3}{a(t)}
\label{eq:expansion1}
\end{equation}
It is obvious that the expansion here is negative, as the
matter is going through a process of continual collapse.
As our main interest is to study the curvature singularity
at $R=0$, we require that there are no shell crossings, that is, 
we have $R'>0$ in the spacetime. This is equivalent to the 
condition that $r\psi'>-1$ as we see from the expression
for $R$ above. We note that the comoving coordinate system 
we have employed here is based
on the congruence of in-falling matter. In this case
then ensuring that there are no shell
crossings would ensure that the coordinate system is valid and 
does not breakdown till the curvature singularity at $R=0$.

It is important to note here that in the  
case of dust, which is a special case of general {\it Type I}
matter fields considered here, the congruence of the curves of 
collapsing matter are {\it geodesics}
and hence in that case no shell crossings would imply there are
no conjugate points in the coordinate system. 
In fact, in this case, we have,
\begin{equation}
\dot\psi = \dot R'/R'
\label{eq:dust}
\end{equation}
so $R'\to 0$ implies $\Theta \to -\infty$, which shows that
the shell-crosses are equivalent to occurrence of conjugate
points in the congruence of geodesics. Again the condition
$R'>0$ then ensures that the coordinate system is valid till
the curvature singularity at $R=0$.
In the general case considered here, the collapsing matter
need not move along geodesics, however, imposing the condition 
as above that there are no shell-crossings ensures that
the co-ordinate system does not break
down till the curvature singularity.

Now for the curvature singulartity at $R=0$, we have 
$\Theta\rightarrow -\infty$ as the singularity acts like a 
{\it sink} for all the curves of the collapsing congruence, and
the volume elements shrink to zero along all the collapsing
trajectories 
\cite{haw}. 
This implies that at the curvature singularity we have $a(t)=0$. 
Further, we note that at all the regular spacetime events $a(t)$
is finite and non-zero. This follows from equation (11) which 
implies that $\dot\psi$ is finite at all regular points because
so are $R$ and $\dot R$. Then from equation (16), since $\nu(r,t)$
has to be regular in the spacetime, it follows that $a(t)$ has 
the above behaviour. This makes physical sense also because
$\Theta$ is the physical parameter characterizing the volume expansion
(or shrinkage) of the collapsing cloud.

Now, let the 
time taken for the central shell at $r=0$
to reach the singularity be denoted by $t_{s_0}$, where the 
singularity curve $t_s(r)$ corresponds to singularity at $R=0$. 
In the present situation, we need to examine if any non-spacelike
trajectories could escape away from the singularity, thus 
making it visible. In other words, we have to find if there are
any future directed non-spacelike curves which reach to faraway 
observers, and in the past which terminate at the singularity.
Suppose now that the singularity curve $t_s(r)$ is an increasing function
with $r$. Consider then the spacelike surface $t= t_{s_0}$.
Any event on this surface with $r>0$ then lies in the spacetime, 
because the singularity $R=0$ for this collapsing shell
is reached at a later time. So this is a regular event at which
$\Theta$ must be finite as shown above. However, this is not possible
as we know that $\Theta(t_{s_0}) = -\infty$. Similar argument 
applies if $t_{s}(r)$ were a decreasing function of $r$. 
It follows that $t_s(r)=t_{s_0}$, which is a constant function,
and we have,
\begin{equation}
\Theta(t_s(r))=\Theta(t_{s_0}) = -\infty
\label{eq:simul}
\end{equation}
This implies that the singularity $t_s(r)$ is {\it simultaneous}.
The singularity curve being constant is a necessary and
sufficient condition for the collapse to be simultaneous.

A simulataneous collapse, however, necessarily must give rise to 
a covered singularity at $R=0$, and there cannot be any 
outgoing future directed non-spacelike geodesics
coming out from the same. Because, if there were any such
outgoing geodesics, given by say $t=t(r)$ in the $(t,r)$ plane, which 
came out from $t=t_s,r=0$, then the time coordinate must increase 
along these paths. This is, however, impossible as there 
is complete collapse at $t=t_{s_0}$, and there is no 
spacetime beyond that. Hence no values $t>t_{s_0}$ are allowed
within the spacetime which does not extend beyond the singularity.
Thus, the collapse gives rise necessarily to a black hole in the 
spacetime 
(see also \cite{gj}).
Similar argument of course applies either to the points on the
singularity curve at $r=0$ or $r>0$.

It follows that in a shear-free collapse of a 
general {\it Type I} matter field, the end state of gravitational
collapse is always necessarily a black hole. What the above shows is that
in this case, all the collapsing shells
reach the singularity simultaneously. 
As a result, the singularity is necessarily covered and is a
black hole.

As we indicated above, we have considered here a sufficiently
general form of matter, and hence our conclusions above on the black hole 
formation as collapse end state are rather generic. To get a better 
insight into how shear 
operates in a dynamically evolving scenario, let us now 
consider
in some detail the case of a collapse evolution where we assume
the matter density to remain homogeneous throughout. We
shall construct below an explicit class of collapse models to 
understand how shear works and affects the collapse.
The choice of a homogeneous density profile of course does not 
mean that we are choosing 
any particular 
form of matter, but it just implies that the density is a 
function of time only, i.e.
\begin{equation}
\rho(r,t)=\rho(t)
\label{eq:rho}
\end{equation}
Now, let us 
choose the class of velocity profiles for the collapsing 
shells as determined by the choice,
\begin{equation}
\n=A(R)
\label{eq:nu1}
\end{equation}
Here the function $A(R)$ is any arbitrary, suitably differentiable 
function of the physical radius $R$ of the cloud, with the initial 
constraint
\begin{equation}
A(R)|_{t=t_i}=\nu_0(r)
\label{eq:A1}
\end{equation}
Again, from the Einstein equation (\ref{eq:ein2}) we get,
\begin{equation}
\nu_0(r)=\int_0^r\left(\frac{2(p_{\T_0}-p_{r_0})}{r(\rho_0+p_{r_0})}
-\frac{p_{r_0}'}{\rho_0+p_{r_0}}\right)dr
\label{eq:nu0}
\end{equation}
Let us now assume that the initial pressures have physically 
reasonable behavior at the center $r=0$, in that the pressure gradients 
vanish, i.e. $\prz'(0)=\ptz'(0)=0$, and also that the difference between 
radial and tangential pressures vanishes at the center, 
i.e. $\prz(0)-\ptz(0)=0$, which ensures the regularity of the initial data 
at the center of the cloud.
Then, from \e(\ref{eq:nu0}), it is evident that $\nu_0(r)$ 
has the form,
\begin{equation}
\nu_0(r)=r^2g(r)
\label{eq:nu0form}
\end{equation} 
where, $g(r)$ is at least a $C^1$ function of $r$ for $r=0$, and at least 
a $C^2$ function for $r>0$.
From equation (\ref{eq:nu0form}) we can now generalize the form 
of $A(R)$ as,
\begin{equation}
A(R)=R^2g_1(R)
\label{eq:A2}
\end{equation}
where $g_1(R)$ is a suitably differentiable function and,
\begin{equation}
g_1(R)|_{t=t_i}=g(r)
\label{eq:A3}
\end{equation}
Now from equations (\ref{eq:rho}) and (\ref{eq:ein1}) we can directly
calculate the function $F$ and $p_r$ as,
\begin{equation}
F=\frac{1}{3}\rho(t)R^3
\label{eq:F}
\end{equation}
and,
\begin{equation}
p_r=-\rho(t)-\frac{1}{3}\dot{\rho(t)}\frac{R}{\dot{R}}
\label{eq:pr}
\end{equation}
Also, using \e (\ref{eq:nu1}) in \e (\ref{eq:ein3}), we have,
\begin{equation}
G(t,r)=b(r)e^{2A}
\label{eq:G}
\end{equation}
Here $b(r)$ is another arbitrary function of the radial coordinate 
$r$. (A comparison
with dust collapse models interprets $b(r)$ as the velocity function for
the shells). We can write $b(r)$ as,
\begin{equation}
b(r)=1+r^2b_0(r)
\label{eq:veldist}
\end{equation}
Thus we see that for an $\epsilon$ ball around the central shell, 
the function $G$ behaves as,
\begin{equation}
G\approx e^{2A}
\label{eq:G1}
\end{equation}
Now using \e (\ref{eq:ein4}) we get
\begin{equation}
\frac{\dot{R}}{R}=-e^{A}\sqrt{2g_1(R)+\frac{1}{3}\rho(t)}
\label{eq:rdot1}
\end{equation}
Hence, it is evident that in the vicinity of the singularity, that is 
in the limit $R\rightarrow 0$ and $\rho\rightarrow\infty$, and close 
to the central shell,
\begin{equation}
\frac{\dot{R}}{R}=f(t)
\label{eq:rdot2}
\end{equation}
Here $f(t)$ is another function of time. Thus from 
\e (\ref{eq:rdot2}) and (\ref{eq:pr}) we see that in the limit 
of approach to the singularity and 
near the central shell, the radial pressure behaves as,
\begin{equation}
p_r=p_r(t)
\label{eq:pr2}
\end{equation}
Again, in the same limit we can write the tangential pressure as,
\begin{equation}
2\pt=RA_{,R}(\rho+\pr)+2\pr\approx2\pr(t)
\label{eq:pt1}
\end{equation}

It is clear therefore from the above that in the case of a 
homogeneous collapse, 
a large class of solutions as given above, and characterized by the 
functions $A(R)$ exists, for 
which both the radial and tangential pressures
homogenize close to the center and in the vicinity of the 
singularity, and hence the collapse becomes necessarily 
shear-free at this limit (note that in spherical symmetry homogeneity, 
through the no energy flux condition, implies vanishing shear). 
The final outcome of such a homogeneous collapse is then
necessarily black hole.

We have thus shown that it is black hole whenever the collapsing matter  
is shear-free or homogeneous. That is, the end product of collapse could 
be different from black hole (i.e. naked singularity) only if the collapsing 
matter is both shearing as well as inhomogeneous. This is an important and a 
fairly general feature because the matter field we have considered 
is general enough. It follows that for a number of classes of general {\it 
Type I} matter fields, the homogeneous collapsing configurations are 
subclasses of shear-free collapse. The point we make here is
a collapse which is homogeneous in density tends to a shear-free
configuration in the limit of approach to the singularity.

An important insight that emerges from our consideration here 
is that there exists a remarkable tying up of shear and inhomogeneity for 
a gravitationally collapsing matter cloud to end up in a naked singularity. 
Intuitively, the physical process that could distort the shape of 
apparent horizon surface so as to expose the singularity is shear. 
That is why shear-free collapse always ends in black hole. Further we have 
also shown that as the singularity is approached, shear could only 
be produced by inhomogeneity in density. Note that this is also the case 
for pressureless dust (and homogeneous pressure) where shear 
is solely generated by density inhomogeneity. What seems to happen 
is that as the singularity is approached pressures turn homogeneous 
and hence ineffective. 
Thus inhomogeneous 
density is necessarily required for shear to be non-zero as the 
singularity is approached. 
Interestingly, the same tying up of inhomogeneity and shear(anisotropy) 
occurs for non-singular cosmological models as was first argued 
by one of us 
\cite {nkd}. 
We hence see that the collapse for a general matter field as
we have considered here generically tends to the outcome which
is observed in the dust case
\cite{jdm}. 
This raises an interesting possibility that generically all
collapse configurations could tend to a dustlike model in the
vicinity of the singularity.

Another interesting feature that seems to follow quite generally and 
generically (where weak energy condition is satisfied with densities and 
pressures being positive), 
when one 
considers the outgoing null geodesics equation, is that as expected 
it is only the central singularity at $r=0$ 
which could be naked and none else for $r>0$. 
To see this, and in order to examine the nature of the central 
singularity at $R=0, r=0$, consider the equation for outgoing radial 
null geodesics, 
\begin{equation}
\frac{dt}{dr}=e^{\psi-\nu}
\end{equation}
One could now write the above equation 
in terms of the variables $(u=r^\alpha,R)$,
where one could choose $\alpha=\frac{5}{3}$, 
and using \e (\ref{eq:ein4})
we get,
\begin{equation}
\frac{dR}{du}=\frac{3}{5}\left(\frac{R}{u}+\frac{\sqrt{v}v'}
{\sqrt{\frac{R}{u}}}\right)\left(\frac{1-\frac{F}{R}}
{\sqrt{G}[\sqrt{G}+\sqrt{H}]}\right)
\label{eq:null3}
\end{equation}
where we have written $R=rv$.
Now if the null geodesics do terminate at the singularity 
in the past with a definite tangent, then at the singularity we have
$\frac{dR}{du}>0$, in the $(u,R)$ plane with a finite value.
It follows that all points $r>0$ on the singularity curve are 
then covered because $F/R \rightarrow\infty$ (when weak energy
condition is satisfied and pressures are positive then $F(r)$ tends
to a finite positive value for any $r>0$ on the singularity curve) 
with $\frac{dR}{du}\rightarrow-\infty$.
This implies that non-central 
singularities will always be covered for a general enough {\it type I} 
matter distribution. The central singularity could however be
visible.

More generally, it can be stated that any spacetime or 
evolving distribution has two dynamically interesting features, namely
the homogeneity and isotropy. Though they are quite different properties 
yet they seem to get quite intimately correlated at the end stage of the 
collapse in the vicinity of the singularity. That is, as the 
singularity is approached one implies the other. The role of shear or 
anisotropy is to make the motion
incoherent. This means the motion is not in the same sense in all 
directions, and the collapse or expansion may proceed at different 
rates in different directions. Intuitively, the shearing force will act 
against the concentration of matter quickly enough in small
enough region, or creates a
distortion, which in turn delays the formation of apparent horizon
or the trapped surfaces. This is how the 
shear, which in the ultimate stage of collapse entirely hinges on density 
inhomogeneity, plays the key role. Thus both shear and inhomogeneous 
density are required if the collapse has to end in a 
naked singularity.

What we have shown here is that for a fairly generic matter 
distribution, the absence of shear would always lead to a black hole. 
Hence in a way we have pointed out and identified here
the physical mechanism that causes naked singularities in collapse. 
In this sense, our results here explain 
in a natural manner why naked singularities do develop as end state of 
collapse, when there are non-vanishing shearing forces and 
inhomogeneities present within the collapsing cloud.\\ 

It is a pleasure to thank Sanjay Jhingan and Louis Witten for discussions.

\end{document}